\documentclass[prb,aps,10pt,twocolumn,groupedaddress,floats,showpacs,final,superscriptaddress,floatfix]{revtex4-2}
\usepackage[makeroom]{cancel}
\usepackage{svg}
\usepackage{amsmath}
\usepackage{amsfonts}
\usepackage{amssymb}
\usepackage{chngcntr}
\usepackage{color}
\usepackage{graphicx}
\usepackage{dcolumn}
\usepackage{bm}
\usepackage{array}
\usepackage{appendix}
\usepackage{xcolor}
\usepackage[
    colorlinks=true,
    citecolor=blue,
    linkcolor=blue,
    urlcolor=blue
]{hyperref}
\definecolor{purple}{rgb}{0.5,0,0.6}

\usepackage{soul}

\begin{document}

\title{Coulomb-mediated interactions of charge-transfer excitons in TMD lateral heterostructures}

\author{Kabyashree~Sonowal}
\affiliation{Department of Physics, Philipps-Universität Marburg, 35037 Marburg, Germany}
\affiliation{mar.quest—Marburg Center for Quantum Materials and Sustainable Technologies, 35032 Marburg, Germany}

\author{Daniel~Erkensten}
\affiliation{Department of Physics, Philipps-Universität Marburg, 35037 Marburg, Germany}
\affiliation{mar.quest—Marburg Center for Quantum Materials and Sustainable Technologies, 35032 Marburg, Germany}

\author{Ermin~Malic}
\affiliation{Department of Physics, Philipps-Universität Marburg, 35037 Marburg, Germany}
\affiliation{mar.quest—Marburg Center for Quantum Materials and Sustainable Technologies, 35032 Marburg, Germany}

\author{Roberto~Rosati}
\affiliation{Department of Physics, Philipps-Universität Marburg, 35037 Marburg, Germany}
\affiliation{mar.quest—Marburg Center for Quantum Materials and Sustainable Technologies, 35032 Marburg, Germany}

\date{\today}

\begin{abstract}
Lateral heterostructures of transition-metal dichalcogenides (TMDs) host spatially separated  charge-transfer (CT) excitons. While analogous to interlayer excitons in vertical TMD heterostructures, these interfacial excitons possess much larger in-plane dipoles of several nanometers and an additional center-of-mass quantization.  Here, we study the mutual interactions between these highly dipolar CT excitons on a microscopic footing. Accounting for the dipolar and quantum exchange interactions, we evaluate the experimentally accessible density-dependent energy renormalization and predict a net energy blueshift of a few meV for bound CT excitons. Interestingly, for small dipole moments, the energy renormalization displays a quadratic dependence with respect to the dipole moment, in contrast to the linear dependence found in vertical TMD heterostructures. We show that spatial energy offset and temperature are the key tuning knobs for controlling the density-dependent excitonic response. Overall, our results contribute to a better microscopic understanding of CT excitons and their interactions in lateral TMD heterostructures.
\end{abstract}

\maketitle

\section{INTRODUCTION}

\begin{figure}
\includegraphics[width=0.9\columnwidth]{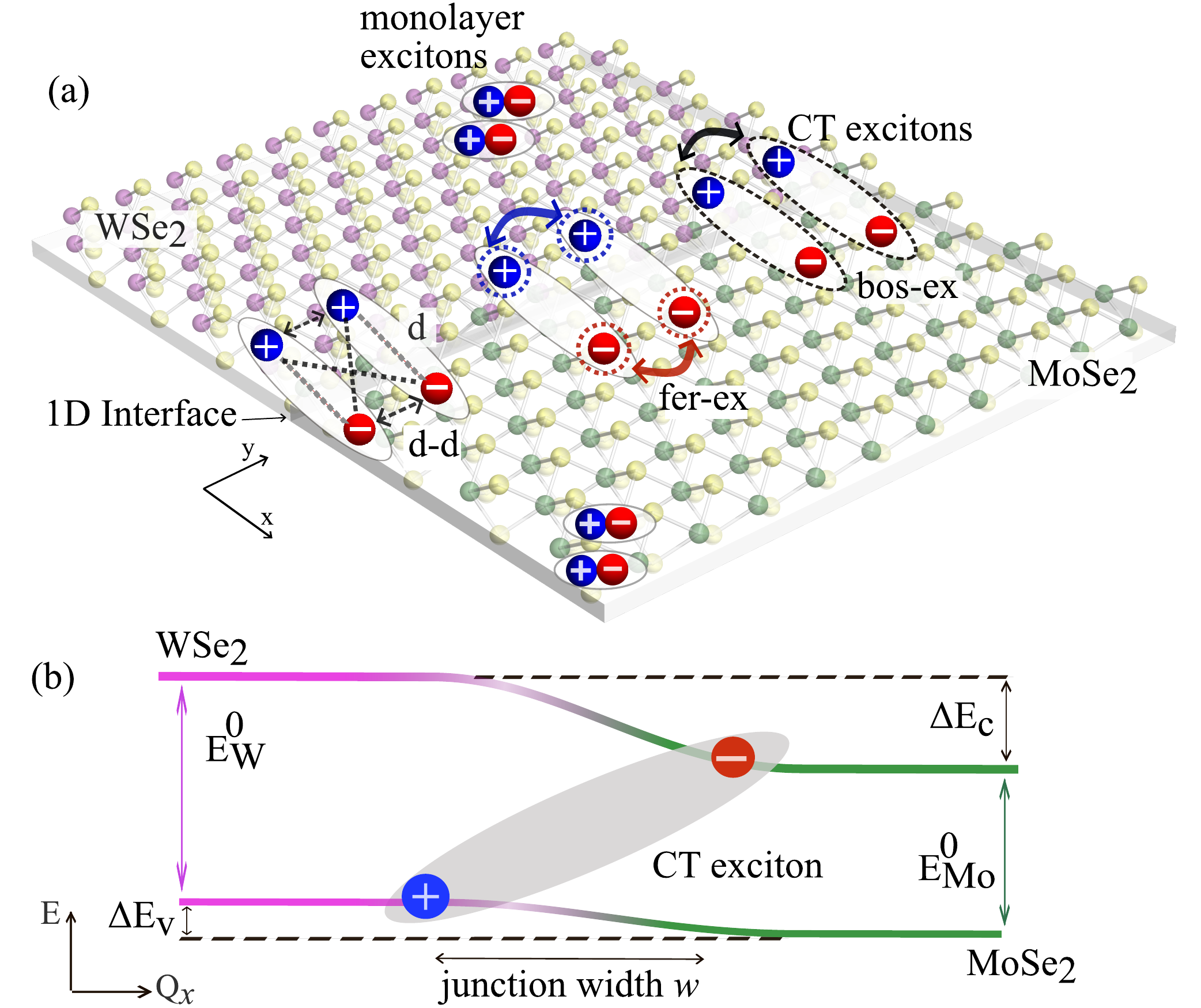} 
\caption{(a) Schematic illustration showing charge transfer (CT) excitons at the interface of the ${\rm MoSe_2-WSe_2}$ lateral heterostructure  as well as regular intralayer excitons far away from the interface. The 1D interface is chosen to be along the y direction. CT excitons interact via dipole-dipole (d-d) interaction, bosonic exchange (bos-ex), and fermionic exchange (fer-ex). (b) Formation of conduction ($\rm \Delta E_{\rm c}$) and valence band offsets ($\rm \Delta E_{\rm v}$) at opposite sides of the interface, where $\rm E_{W}^0$ and $\rm E_{Mo}^0$ are intrinsic single-particle band gaps of the constituent TMD monolayers.}
\label{Fig1}
\end{figure}

Lateral heterostructures (LHs) of transition metal dichalcogenides (TMDs) \cite{Duan14,Gong14,Huang14,Li15,Heo15,Zhang18,sahoo2018one,Fali21,Zhang22,Sousa25,Kundu24,Shradha26,Kundu26,Kaur26,Huang26} consist of two or more monolayers that are connected by in-plane covalent bonds (Fig.~\ref{Fig1}(a)).  At the junction formed between different TMD monolayers, the presence of a type-II band alignment~\cite{robertson2016bandengineering} facilitates the formation of charge-transfer (CT) excitons (Fig.~\ref{Fig1}(b)). They consist of  spatially separated electrons and holes  bound by the Coulomb interaction across opposite sides of the junction~\cite{lau2018interface, rosati2023interface,Yuan23,rosati2025impact,Durnev25,vandoolaeghe2025,durnev2026tunable}.  
Experimental evidence of their existence at the interface of  ${\rm MoSe_2 - WSe_2}$ and WS$_{1.16}-$WSe$_{0.84}$ LHs has recently been reported \cite{rosati2023interface,yuan2023strong,vandoolaeghe2025}. 
This observation was made possible due to atomically sharp 1D interfaces on the order of the exciton Bohr radius \cite{rosati2023interface}
, as allowed by recent advances in bottom-up approaches for fabricating LHs. \cite{Najafidehaghani21,Li15,Xie18,Ichinose22,sahoo2018one}. 

LHs have recently emerged as a unique platform for studying exciton optics, dynamics, and transport~\cite{Saireview2025,Rosati2025chapter,Smirnov2025intervalley,durnev2026tunable,Beret22,rosati2025impact}. They host a spatially varying exciton energy landscape created by the energetic offset between intralayer excitons of the constituent TMD monolayers at the opposite sides of the interface, facilitating  unidirectional exciton transport across the interface~\cite{Beret22,rosati2025impact,lamsaadi2023kapitza,Bellus18,Shimasaki22,Kundu24,Lamsaadi25,rosati2026exciton}. Exciton propagation is crucially affected by exciton capture at the interface~\cite{rosati2025impact} resulting in a high CT exciton density, which can drive a non-linear enhancement of exciton diffusion along the interface \cite{yuan2023strong}. The exciton landscape in LHs can be tuned by varying the energetic band offset, controlling the interface width, and by dielectric engineering~\cite{rosati2023interface}. This allows us to tune the spatial dipole moment and exciton binding energies, making LHs a potential material platform for tunable dipolar interactions \cite{sun2024dipolar}. 

Dipolar interactions have been shown to play a key role in driving quantum many-body phenomena in vertically stacked TMD heterostructures (VHs)~\cite{sun2024dipolar}. They host interlayer excitons with finite out-of-plane dipole moment resulting in strong repulsive inter-excitonic interactions and efficient transport of interlayer excitons ~\cite{erkensten2021exciton,erkensten2022microscopic,sun2022excitonic,Hagel2021,perea2022exciton,schmitt2022formation}. The spatial dipole in VHs is geometrically constrained and can only be varied by adding dielectric spacers~\cite{erkensten2023electrically}. In contrast, the in-plane dipole in LHs can extend up to tens of nanometers~\cite{rosati2023interface}. Moreover, CT excitons exhibit an additional center-of-mass  (COM) quantization at the interface, in contrast to monolayer or interlayer excitons, which are distributed over the 2D sample. Thus, CT excitons are crucial to investigate tunable dipolar interactions with  one-dimensional confinement~\cite{Lau18,rosati2023interface}. While such interactions have been intensively studied in VHs~\cite{erkensten2022microscopic,erkensten2021exciton,steinhoff2024exciton}, a consistent microscopic understanding of inter-excitonic interactions in LHs is still lacking.

In this theoretical work, we study Coulomb-mediated interactions \cite{erkensten2022microscopic,ciuti1998role,Lengers19,steinhoff2024exciton,Grisard24,shahnazaryan2017exciton, koenig16,mittenzwey2026coulomb,Mittenzwey26} between CT excitons on a microscopic footing and analyze density-dependent energy renormalization at a Hartree-Fock level~\cite{Koch2004}. We predict a density-dependent net blueshift resulting from the competition between repulsive dipolar and bosonic exchange interactions, and the attractive fermionic exchange interaction (Fig.~\ref{Fig1}(a)). In the experimentally relevant case of hBN-encapsulated MoSe$_2$-WSe$_2$ LHs, we find that the net blueshift of CT excitons can reach up to tens of meV. Interestingly, the predicted energy renormalization is comparable to the observed blueshift in VHs at similar densities, despite the much smaller dipole moment of interlayer excitons~\cite{erkensten2022microscopic}.  For small dipoles, we find a quadratic  dependence of the energy renormalization as a function of the dipole moment, clearly deviating from the linear dependence  in VHs~\cite{erkensten2022microscopic}. We find that the band offset plays a crucial role in determining the nature of energy shifts. Furthermore, varying the temperature leads to non-monotonic variations in the energy shift arising from the competition between different temperature-dependent mechanisms.


\section{Theoretical model}\label{theory}

In order to analyze exciton-exciton interactions in LHs, we employ a theoretical framework based on an equation-of-motion approach~\cite{katsch2018theory,erkensten2022microscopic} and combine it with a microscopic evaluation of CT exciton wavefunctions and energies. We introduce the microscopic polarisation $\hat{P}^{\dag}_{\bf k,\bf k'} = \hat{c}^{\dag}_{\bf k}\hat{v}_{\bf k'}$, 
where $\hat c^{(\dagger)}_{\textbf{k}}$ and $\hat v^{(\dagger)}_{\textbf{k}}$ 
are the annihilation (creation) operators of an electron with the momentum $\textbf{k} = (k_x,k_y)$ in the conduction and valence band, respectively. We investigate the dynamics of its expectation value $\langle\hat{P}^{\dag}_{\bf k,\bf k'} \rangle$ 
using the Heisenberg equation of motion with the many-particle Hamiltonian including single-particle energies and the Coulomb interaction in the electron-hole basis. The resulting equation includes expectation values of the product of four electronic operators, which we expand into pair operators. Then, we transform the equation into the excitonic picture, and 
focus on the microscopic polarization induced by CT excitons  $\hat{P}^{\dag}_{\bf k,\bf k'}\approx \hat{X}^{\dag}_{CT, k_y-k_y'} \Psi^{\ast}(k_x-k_x',\beta {\bf k} + \alpha {\bf k'})$, where we here restricted to the energetically lowest CT exciton states. Furthermore, we introduce the CT exciton operator $\hat{X}^{\dag}_{CT,Q_y}=\sum_{Q_x,\bm q}{\Psi}(Q_x,\bm q)\hat{v}_{\bm q-\beta \bm Q}\hat{c}_{\bm q + \alpha \bm Q}^{\dag}$, where $Q_x$ and $Q_y$ are COM momenta across and along the interface (Fig.~\ref{Fig1}), respectively.  Here, $\alpha=m_e/(m_e+m_h)$ and $\beta=m_h/(m_e + m_h)$ are obtained from the electron and hole effective masses $m_e$ and $m_h$, respectively. 
Here, $\Psi(Q_x,\bm q)$ is the Fourier transform of the CT exciton wavefunction ${\Psi}(R_x,\bm r)={\psi}(R_x){\phi}^{R_x}(\bm r)$, where $R_x$ denotes the COM coordinate across the interface and ${\bm r} = (x,y)$ is the relative space coordinate. We note that owing to the asymmetric band alignment across the interface (see fig.~\ref{Fig1}(b)), the COM wavefunction in $R_x$ becomes non-trivial due to a finite localization, whereas the $R_y$ component along the interface remains unaffected and retains its plane-wave character. Furthermore, ${\psi}(R_x)$ and ${\phi}^{R_x}(\bm r)$ represent wavefunctions obtained by numerically solving the corresponding  Schrödinger equations for COM and relative motion, respectively, for different valence band offsets $\rm \Delta E_{v}$ (Fig.~\ref{Fig1}(b))~\cite{Lau18,rosati2023interface}. Then, we employ a mean-value approximation such that the COM and the relative motion of CT excitons can be  decoupled, i.e.,  $\Psi(Q_x,{\bm q}) = \psi(Q_x)\bar\phi({\bm q})$, where $\bar\phi({\bm q})$ is the Fourier transform of the full wavefunction averaged over COM position $R_X$(see Supplementary Information (SI) for additional details).

Investigating the dynamics of the microscopic polarization $P_{Q_y} = \langle \hat{X}^{\dag}_{CT,Q_y} \rangle$ of the bright 1s exciton ($Q_y=0$) and focusing on the Coulomb-mediated energy renormalization, we obtain $-i\hbar \partial_t P_0|_{\mathrm{ren}}=\Delta E P_0$. Here, $\Delta E$ is the density-dependent energy shift for excitons in the lightcone, which is given by
\begin{align}\label{Eshift}
        \Delta E = n_{1D} L_y \Big( g^{\rm d-d} + g^{\rm bos-ex} + g^{\rm fer-ex}\Big) \ , 
\end{align}
where $n_{1D}= \sum_{Q_y}{N_{Q_y}}/{L_y}$ is the one-dimensional CT exciton density with $L_y$ as the length along the interface, and $N_{Q_y}=\langle \hat{X}^{\dag}_{Q_y,CT}\hat{X}_{Q_y,CT} \rangle$ denoting the CT-exciton occupation. The corresponding energy renormalization for excitons with finite COM momentum $Q_y$ can be found in the SI. While excitons can be trapped at the interface resulting in a more involved dynamics of CT excitons \cite{rosati2025impact,rosati2026exciton}, we focus here on  the equilibrium situation and consider pristine and unscreened LHs. Dynamic screening effects are expected to quantitatively reduce the overall energy shift, as shown for TMD monolayers and VHs~\cite{erkensten2022microscopic,steinhoff2024exciton}.

The first term in Eq. \eqref{Eshift}, $g^{\rm d-d} = W^{\rm direct}_0$ represents the classical dipole-dipole repulsion. The second term reflects the bosonic exchange of individual excitons and is given by the thermal average of the momentum-dependent direct element $g^{\rm bos-ex} =\sum_{Q_{y}}N_{Q_{y}}W^{\rm direct}_{Q_y}/\sum_{Q_y} N_{Q_y}$. 
The third term reflects the fermionic character of  excitons and is given by the carrier exchange term  $g^{\rm fer-ex}  =  \sum_{Q_y} N_{Q_y} ~(W^{\rm fer-ex}_{Q_y,0} + W^{\rm fer-ex}_{Q_y,-Q_y})/\sum_{Q_y} N_{Q_y}$. The interaction matrix elements  $W^{\rm direct}$ and $W^{\rm fer-ex}$ correspond to the direct (Hartree)  and the Fock term,  respectively. They consist of the product of COM and relative momentum wavefunctions $\psi(Q_x)$ and $\bar{\phi}({\bm q})$, respectively, together with the Coulomb interaction $V_{\bm q}$ given by the Keldysh potential \cite{Keldysh79,Rytova67} (see SI for more details). As both the dipole-dipole interaction and bosonic exchange originate from the direct term, we summarize them to one contribution and refer to it as the direct term.  Note that in the limit of vanishing temperature, $T\rightarrow 0~$K, both contributions coincide.

The complete evaluation of the interaction matrix elements is involved due to the presence of multiple nested momentum summations. Hence,    we analytically model CT exciton wavefunctions $\psi(Q_x)$ and $\bar{\phi}({\bm q})$ as Gaussian functions characterized by three microscopically calculated quantities: (i) $\Delta_X$ denoting the spatial width or confinement length of the COM wavefunction, (ii) $\Delta_r$ representing the spatial extent or average squared width of the CT exciton wavefunction in relative coordinates and (iii) the dipole moment ${\bf d} = {\rm d}\, \hat{x}$.
The values of $\Delta_X, \Delta_r$ and $\rm d$ are obtained from the numerical solution of the Schrodinger equations as detailed in the SI. Inserting these values into the analytically modeled Gaussian functions results in a wavefunction $\Psi(Q_x,{\bm q})$ that is in good agreement with the fully microscopically calculated wavefunction ~\cite{rosati2023interface}. Note that in the limit $\Delta_X \to \infty$, the COM wavefunction of CT excitons becomes delocalized as in the case of VHs and TMD monolayers. In this limit, taking $T \to 0$ K,  setting ${\rm d}=0$, and substituting $\Delta_r$ with the excitonic Bohr radius, we obtain the energy shifts of intralayer excitons in TMD monolayers (see SI). Moreover, in the same limit but for a fixed finite dipole $\rm {d}\neq 0$, and assuming a layer-dependent Coulomb potential~\cite{ovesen2019interlayer},  the interaction matrix elements for interlayer excitons in VHs can be reproduced~\cite{erkensten2022microscopic}.

\section{results}
\subsection{Density-dependent energy renormalization}\label{sbI}
\begin{figure}[t!]
\includegraphics[width=0.9\columnwidth]{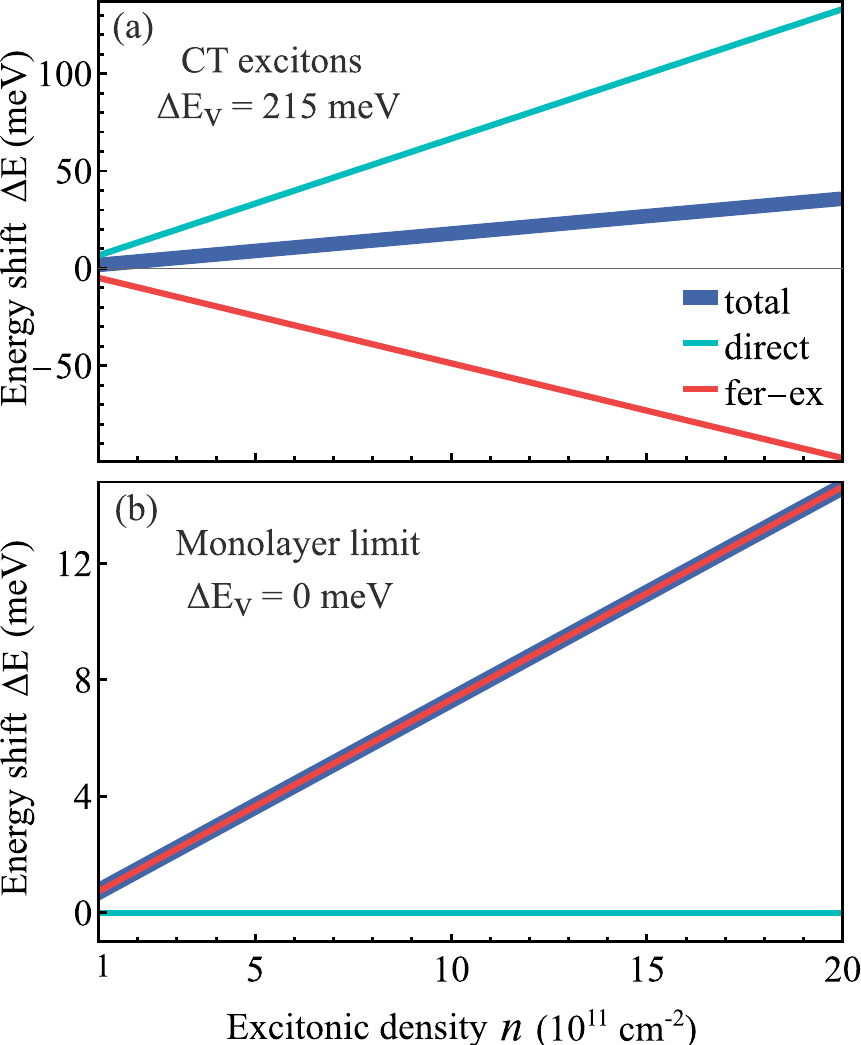} 
\caption{Density-dependent energy renormalization as a function of 2D exciton densities  at T = 0 K for two different band offsets: (a) $\rm \Delta E_{\rm v} =$ 215 meV, where bound CT excitons are present, and (b) $\rm \Delta E_{\rm v} =$ 0 meV corresponding to the monolayer exciton limit. Note that the direct term comprises the dipole-dipole interaction as well as the bosonic exchange, which contribute equally at vanishing temperature. }
\label{Fig2}
\end{figure}

Firstly, we study the density-dependent energy renormalization in the hBN-encapsulated LH at $T=0~$K. For this purpose, we numerically evaluate Eq.~\eqref{Eshift} for a range of CT exciton densities for the experimentally relevant case of interface width of $w = 2.4$ nm, which allows observation of CT excitons in $\rm MoSe_2-WSe_2$ LH. ~\cite{rosati2023interface}.  In Fig.~\ref{Fig2}, we show the resulting density-dependent energy shifts for two different values of energetic band offsets: (a) $\rm \Delta E_{\rm v} = 215~meV$  corresponding to $\rm MoSe_2-WSe_2$ LH \cite{rosati2023interface} and (b)  $\rm \Delta E_v = 0~meV$ corresponding to the monolayer limit.  To compare the energy shift of CT excitons trapped along the 1D interface with that of intralayer excitons in 2D monolayers, we express the 1D excitonic density $n_{1D}$ in terms of the corresponding 2D densities  $n$ at the interface $n\equiv n_{1D} |{\psi}(R_x=0)|^2$, see the SI for more details.
 We consider $n\sim 10^{11}-10^{12} {\rm cm}^{-2}$  corresponding to excitonic densities clearly  below the Mott transition in TMD monolayers and vertical TMD heterostructures~\cite{steinhoff2017exciton,Siday22}. 
  We find that for CT excitons   the direct term comprising of the dipolar repulsion and bosonic exchange leads to a blueshift of the order of tens of meV, see Fig.~\ref{Fig2}(a). However, this is largely compensated for by a redshift  arising from fermionic exchange interactions. This competition between the two contributions of the exciton-exciton interactions results in a net blueshift, which increases linearly from 2 to $35$ meV for excitonic densities in the range $\sim 10^{11}-20 \times 10^{11}~{\rm cm}^{-2}$. 

\begin{figure}[t]
\includegraphics[width=\columnwidth]{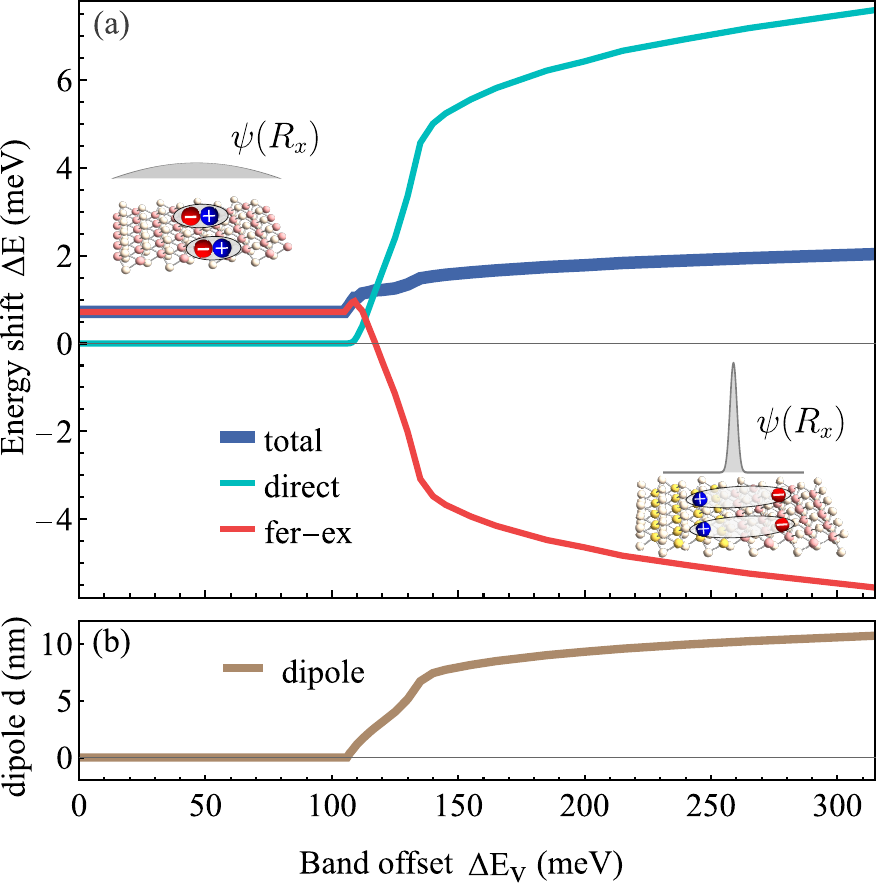} 
\caption{(a) Exciton energy shifts in  MoSe$_2$-WSe$_2$ LHs as function of the energy band offset. The schematics in the insets show intralayer excitons at low band offsets (with a delocalised COM wavefunction) and  bound CT excitons at the interface (with a confined COM wavefunction) at higher band offsets. (b) Dipole moment of CT excitons as a function of the band offset. We consider T = 0 K and the exciton density is fixed to $n=10^{11}~{\rm cm^{-2}}$.  }
\label{Fig3}
\end{figure}

 In the monolayer limit, the exciton ground state has no dipole moment. As a consequence, the contribution of the dipole-dipole interaction and bosonic exchange is zero, cf. cyan line in  Fig.~\ref{Fig2}(b). Nevertheless, we still find a net blueshift for band offsets smaller than 100 meV. This is induced by the  fermionic exchange, which transitions from attractive to repulsive values, when going from dipolar to monolayer excitons, as already shown for TMD monolayers, VHs~\cite{erkensten2022microscopic}, and coupled quantum wells \cite{ciuti1998role,tassone1999exciton}.
In LHs, this transition can be studied by varying the  band offset (see Fig.~\ref{Fig3}), which can be realized by engineering different LHs \cite{rosati2023interface,yuan2023strong}.
The band offset has multiple effects on the ground state, in particular, resulting in a different exciton confinement and dipole moment.  While intralayer excitons dominate for small $\rm \Delta E_v$, band offsets larger than 100 meV allow the formation of  CT excitons \cite{rosati2023interface}.
 As sketched in the inset of Fig.~\ref{Fig3}(a), this results in a confinement of the COM wavefunction $\Psi(R_x)$.   Moreover, the dipole moment increases with the  band offset, extending up to 10 nms for larger band offsets of around 300 meV (see Fig.~\ref{Fig3}(b)).
 
 In Fig.~\ref{Fig3}(a), we analyze the behaviour of the Coulomb-induced energy renormalization as a function of the band offset $\rm \Delta E_v$ for a finite density $n=10^{11}~{\rm cm^{-2}}$. We find that a finite dipole moment for $\rm \Delta E_v>100$\,meV results in the activation of the dipole-dipole repulsion, which abruptly increases from 0 to 5 meV (cyan line in Fig.~\ref{Fig3}(a)). At the same time, the fermionic exchange has an opposite behaviour and it significantly decreases as a function of the band offset(red line in Fig.~\ref{Fig3}(a)).  The cancellation between these two opposite contributions results in an overall net blueshift that  is three times smaller than the shift induced only by the dipole-dipole interaction. Furthermore, we find a piecewise behavior that results from the activation of the two interactions at different band offsets: At high band offsets, the overall blueshift becomes larger reflecting the increased dipole moment and the formation of CT excitons.  At band offsets lower than 100 meV, the dipole moment vanishes and the COM wavefunction becomes fully delocalized. In this regime, our results recover the energy shifts for intralayer excitons in TMD monolayers, where the dipolar repulsion is zero and we find a net blueshift due to the fermionic exchange~\cite{erkensten2022microscopic}.

\subsection{Quadratic dependence on dipole moment}\label{IIIB}
\begin{figure}[t]
\includegraphics[width=0.98\columnwidth]{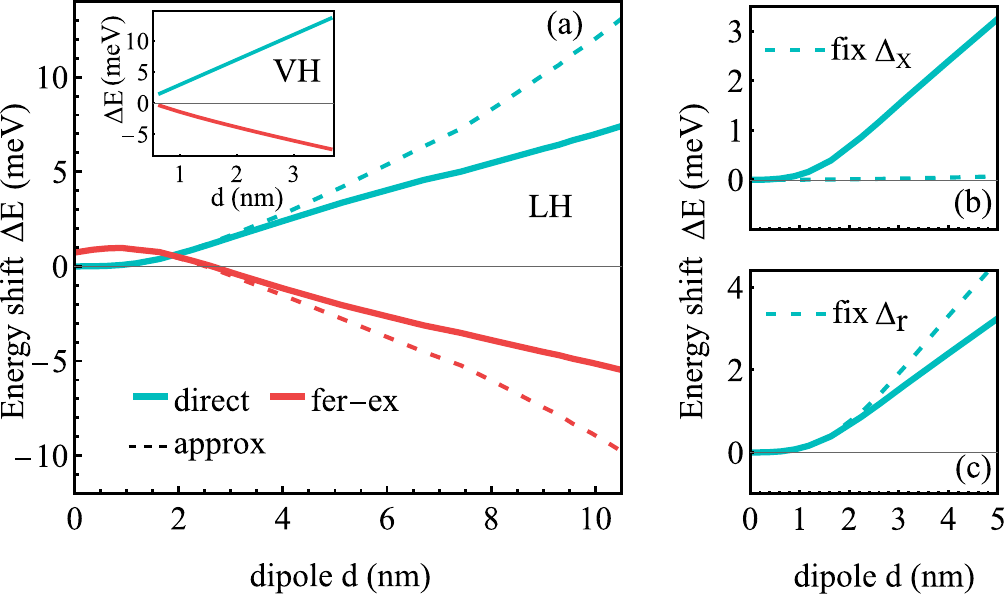} 
\caption{(a) Dipole dependence of the Coulomb-induced energy renormalization at T=0 K for a fixed  density of $n=10^{11}~{\rm cm^{-2}}$ of CT excitons in LHs (main plot) and interlayer excitons  in VHs (inset). The dipole moment of interlayer excitons is tuned by adjusting the hBN spacer thickness of the MoSe$_2$-hBN-WSe$_2$ heterostructure. The dashed lines denote the energy shifts within the small-dipole approximation. Modification of energy shift in the hytpothetical cases of (b) delocalized COM ($\Delta_X = \Delta_X|_{\rm \Delta E_v = 0})$ and (c)  tightly bound CT excitons in relative space ($\Delta_r = \Delta_r|_{\rm \Delta E_v = 0})$.}
\label{Fig4}
\end{figure}

We now analyze the dipole dependence of the density-dependent energy shifts obtained for different energy band offsets. Here, we stress that a variation of the band offset does not only result in a change of the dipole moment $\rm d$, but also in a change of the spatial widths $\Delta_X$  and $\Delta_r$ of the COM  and relative  wavefunctions of CT excitons, respectively. In Fig.\ref{Fig4}(a), we show the energy shift as a function of the dipole moment demonstrating a non-linear dependence, contrary to the linear scaling found for indirect excitons in VHs and coupled quantum wells~\cite{erkensten2022microscopic,ciuti1998role}. To obtain a better understanding of this unexpected behavior, we simplify Eq.~\ref{Eshift} describing the energy shift due to exciton-exciton interactions. Assuming the small-dipole approximation ($q_x{\rm d} \ll 1$), the energy shifts stemming from the direct Coulomb term can be analytically evaluated as $\Delta E_{\rm direct} \approx \sum_{q_x}  V_{q_x} ( a(q_x) + q_x^2 d^2 b(q_x))$
with $a(q_x)$ and $b(q_x)$ being determined by band offset dependent form factors, see the SI for more details.
This indicates that for small dipoles, the energy shifts increase quadratically with the dipole moment and can be clearly separated into dipole-dependent and dipole-independent contributions. In Fig.\ref{Fig4}(a), the dashed lines represent the shifts within the small-dipole approximation, which agree well with the full simulation up to dipoles of approximately 5 nm. The fermionic exchange can also be expressed in terms of zero and quadratic finite dipole contributions in the small-dipole approximation  
with an opposite sign (cf. red line in Fig.\ref{Fig4}(a) and SI). This sign change of the energy shift indicates a blue/red-shift crossover of the fermionic exchange contribution.

Now, we compare the dipole-dependent contributions of  CT excitons in LHs to the corresponding contributions  of interlayer excitons in VHs (inset of Fig.~\ref{Fig4}(a)). The latter is obtained by calculating the energy shift for varying dipole moments that can be tuned by including dielectric hBN spacers between constituent monolayers~\cite{erkensten2023electrically}. The comparison with LHs clearly indicates the contrasting quadratic and linear dipole dependence. Notably, LHs exhibit total energy shifts comparable to those in VHs, even though the dipole moment in VHs is typically smaller. This points to the additional role of $\Delta_X$ and $\Delta_r$ for understanding the energy shift. For increasing band offsets, not only the dipole moment becomes higher, but also the COM wavefunction becomes more confined, i.e., $\Delta_X$ varies from large values of $\sim$ 400 nm (subjected to finite boundary conditions) to finite small values up to around 2 nm. Furthermore,  CT excitons become less bound leading to an increase in $\Delta_r$ at higher band offsets. To analyze the effect of COM confinement in CT excitons  we consider the hypothetical case of delocalised COM in Fig.~\ref{Fig4}(b). Here $\Delta_X$ remains unchanged and has a fixed value $\Delta_X=$ 402 nm obtained with periodic boundary conditions at ${\rm \Delta E_v = 0}$ meV, while d and $\Delta_r$ are varied as a function of the energy offset. We observe that in this limit the direct interaction vanishes even for finite dipoles, indicating that the confinement of COM wavefunction of CT excitons is crucial  to understand the predicted behaviour of the energy shift. We perform a similar analysis to study the role of $\Delta_r$. Here, we keep $\Delta_r$ constant corresponding to the value at zero energy offset and vary $\Delta_X$ and $\rm d$ (Fig.~\ref{Fig4}(c)). We take $\Delta_r$ = 2.1 nm resulting from our numerical calculations at $\rm \Delta E_v = 0$ meV, which is in close agreement with the excitonic Bohr radius reported in monolayers~\cite{Zipfel18}. We find that the effect of $\Delta_r$ on the energy shift is less important than $\Delta_X$. Its contribution becomes relevant only at larger dipole moments.

\subsection{Temperature control of energy shifts}

\begin{figure}[t]
\includegraphics[width=0.9\columnwidth]{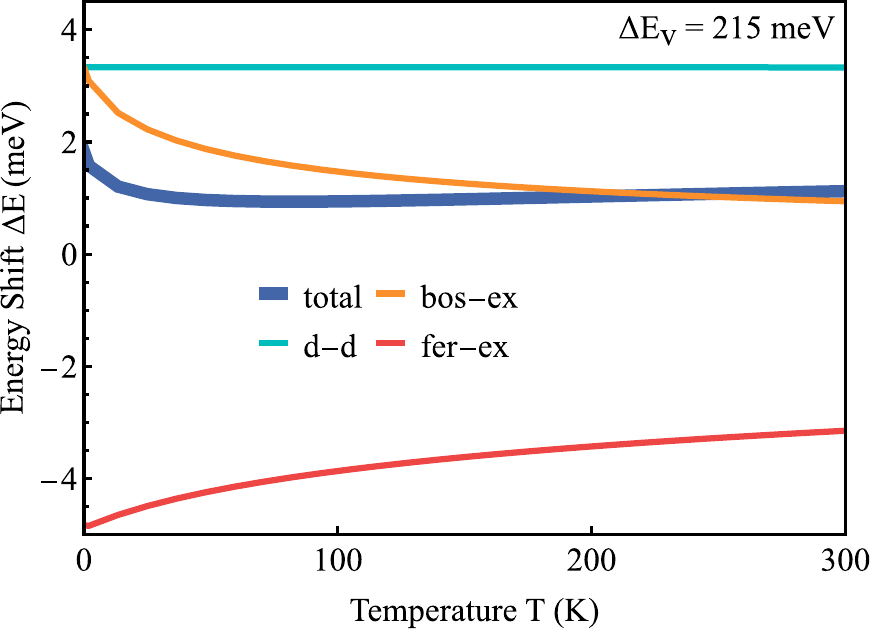} 
\vspace{0.2cm}
\caption{Energy shift as a function of temperature for CT excitons at $\rm \Delta E_v = 215~meV$ and at a fixed density of $n=10^{11} {\rm cm}^{-2}$.  
}
\label{Fig5}
\end{figure}

Finally, we investigate the temperature dependence of Coulomb-mediated interactions of CT excitons. At finite temperatures, the momentum-dependent CT exciton population enters the interaction matrix elements in Eq.~\eqref{Eshift} resulting in a strongly temperature-dependent energy renormalization of CT excitons. For the studied range of exciton densities around $n \approx 10^{11}~\rm{cm^{-2}}$, we consider the 1D Boltzmann distribution  $N(Q_{y}) = \sqrt{2\pi} n_{1D}\lambda_T e^{-\lambda_T^2 Q_y^2/2}$, where  $\lambda_T=\hbar/\sqrt{m k_B T}$ represents the thermal de-Broglie wavelength. We now analyze the effect of  finite temperature on different contributions to the energy shift in Fig.~\ref{Fig5}. Here, we focus on the energy offset of  $\rm \Delta E_{\rm v} =$ 215 meV and find that with increasing temperature, the direct term now constitutes distinct dipole-dipole and bosonic exchange contributions, which coincides at $T=0~$K (see cyan and orange lines in Fig.~\ref{Fig5}). The overall interaction is repulsive. However, the net repulsion becomes less efficient for temperatures increasing from 0\,K to approximately 80\,K followed by a small increase in the energy shift when approaching room temperature (blue line). This non-monotonic behavior is due to the competition of different temperature-dependent mechanisms: While the classical dipole-dipole interaction is independent of temperature,  quantum exchange interactions are weakened at higher temperatures. At $T \approx 0$\,K, the bosonic exchange is as repulsive as the dipole-dipole coupling, both being half of the direct interaction shown in Fig. \ref{Fig3}(a). Then, the repulsive bosonic exchange loses 50\% of its efficiency already at $ T=~$80\,K, while the attractive fermionic exchange contribution decreases only by less than 20\% in the same temperature range. This explains the initial reduction of the energy shift at smaller temperatures. In contrast, at high $T$ the repulsive bosonic exchange is already small, and the decrease of the attractive fermionic exchange gives rise to a slight increase in the overall energy renormalization.

\section{Conclusion}
In this work, we present a microscopic study of Coulomb-mediated interactions between CT excitons in lateral TMD heterostructures. We analyze the resulting energy renormalization over a range of CT exciton densities, temperatures, and energy band offsets at the interface. We find a net blueshift of a few meVs that scales linearly with the exciton density and demonstrates a quadratic dependence on the dipole moment - in contrast to the linear behaviour in vertical TMD heterostructures.  We find that the temperature acts as a tuning knob to control the energy renormalization of bound CT excitons by introducing non-monotonic variations. Our study elaborates on the distinctive role played by the characteristic length scales of  CT excitons including the dipole length and the spatial width of center-of-mass and relative wavefunctions governing exciton-exciton interactions and determining the energy shift.  The comparison of inter-excitonic interactions of CT excitons in lateral TMD heterostructures to  interlayer excitons in vertical TMD heterostructures, and intralayer excitons in TMD monolayers sheds light on the intriguing exciton physics in these technologically promising nanomaterials. 

\section{Acknowledgements}
We acknowledge financial support by the Deutsche Forschungsgemeinschaft (DFG) via the regular projects 535253440 and 542873285. K.S acknowledges support from the Alexander Von Humboldt Foundation. We thank I. Paradisanos for discussions.

\bibliographystyle{apsrev4-2}
\bibliography{bibliography,rosatiBib_short}
\end{document}